%% file: main.tex
\documentclass[runningheads]{llncs}

\usepackage{eccv}

\usepackage{eccvabbrv}

\usepackage{graphicx}
\usepackage{booktabs}
\usepackage{multirow} 
\usepackage{subcaption}
\usepackage{amssymb}
\usepackage[accsupp]{axessibility}  

\usepackage{hyperref}
\usepackage[capitalize]{cleveref}

\usepackage{orcidlink}

\begin{document}

\title{Unveiling Visual Biases in Audio-Visual Localization Benchmarks} 
\titlerunning{Unveiling Visual Biases in AVSL Benchmarks}

\author{Liangyu Chen~~~~~
Zihao Yue~~~~~
Boshen Xu~~~~~
Qin Jin\thanks{Corresponding author.}}

\authorrunning{L.~Chen et al.}

\institute{Renmin University of China\\
\email{\{liangyuchen,yzihao,boshenx,qjin\}@ruc.edu.cn}}

\maketitle


\input{Sections/00_Abstract}
\input{Sections/01_Introduction}
\input{Sections/02_RelatedWork}
\input{Sections/03_VGG-SS}
\input{Sections/04_Epic-SS}
\input{Sections/05_Discussion}
\input{Sections/06_Conclusion}


%
%
\bibliographystyle{splncs04}
\bibliography{main}
\end{document}

%% file: Sections/00_Abstract.tex
\begin{abstract}
Audio-Visual Source Localization (AVSL) aims to localize the source of sound within a video. 
In this paper, we identify a significant issue in existing benchmarks: the sounding objects are often easily recognized based solely on visual cues, which we refer to as visual bias. Such biases hinder these benchmarks from effectively evaluating AVSL models. To further validate our hypothesis regarding visual biases, we examine two representative AVSL benchmarks, VGG-SS and Epic-Sounding-Object, where the vision-only models outperform all audio-visual baselines. Our findings suggest that existing AVSL benchmarks need further refinement to facilitate audio-visual learning.
\keywords{Audio-visual learning \and Audio-visual sound localization \and Benchmark}
\end{abstract}

%% file: Sections/01_Introduction.tex
\section{Introduction}
\label{sec:intro}

{Audio-Visual Source Localization (AVSL)~\cite{avsurvey,look,AVlearning2} aims to ground the sounding objects in the visual scene. 
This task assesses an AI system's ability to correlate sound with corresponding visuals, potentially benefiting various downstream applications, such as VR/AR. 
As one of the key tasks of audio-visual learning, AVSL has been widely investigated over time. 
Research spans across semi-supervised~\cite{learningtolocalsound,dmt}, weakly-supervised~\cite{AVGN,STM}, and self-supervised settings~\cite{hardway1,hardway2,easyway,pixelsthatsound,fnac,SLAVC,SSL-TIE,STM,lin2023unsupervised}, covering scenarios including both first-person~\cite{egoloc} and third-person visual views~\cite{learningtolocalsound}.}

\begin{figure}[tb]
  \centering
  \includegraphics[width=0.95\linewidth]{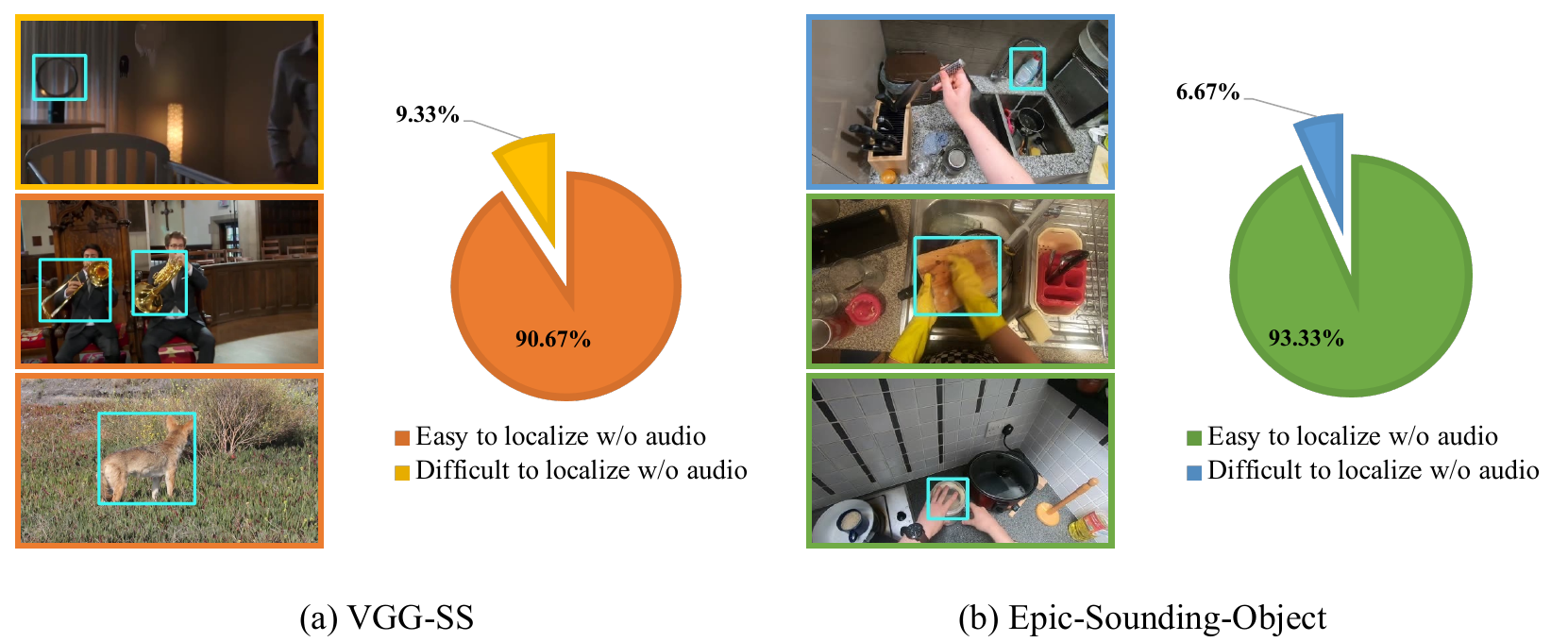}
  \caption{\textbf{Illustration of visual biases in AVSL benchmarks.} Over 90\% sound sources can be accurately identified using only visual information in each randomly sampled set of 300 videos from VGG-SS and Epic-Sounding-Object, respectively.}
  \label{fig:human_eval}
\end{figure}

{However, despite the growing number of model algorithms developed for the AVSL task~\cite{AVGN,mixture,easyway,SSPL,SLAVC,fnac,pixelsthatsound,SSL-TIE,hardway1,hardway2}, the benchmarks commonly used for performance validation have not been thoroughly examined. Beyond issues of the limited data scale and domain diversity,  this paper addresses a fundamental question: \textit{do we really need audio information for sounding object localization} in these benchmarks? For example, when a video shows someone playing an instrument, the source of the sounding object can often be easily deduced from the visual context alone. Similarly, in scenes where someone is cooking in the kitchen, the sounds are likely to originate from human activities such as washing and chopping vegetables. Through our extensive observation over two representative AVSL benchmarks, VGG-SS\cite{hardway1,hardway2} and Epic-Sounding-Object\cite{egoloc}, we surprisingly find that in nearly 90\% cases, the sound source can be accurately inferred using only visual information, as shown in \cref{fig:human_eval}. Such biases towards visually recognizable sounding objects, referred to as visual bias in this paper, hinder existing benchmarks from providing useful feedback for the development of audio-visual models.}


To further confirm the visual biases widely present in existing benchmarks\cite{learningtolocalsound,egoloc}, we investigate how well vision-only models (i.e., models with only visual input) perform on the AVSL task. For the third-person view VGG-SS~\cite{learningtolocalsound} benchmark, to handle the diverse in-the-wild test data, we validate with Large Vision-Language Models (LVLMs) that possess general visual understanding and commonsense reasoning capabilities, particularly, MiniGPT-v2~\cite{minigptv2}. For Epic-Sounding-Object~\cite{egoloc}, where the data is primarily about human hand activities in the kitchen, we assume that simply identifying hand-object interactions (HOI) can often correctly pinpoint the sounding area, and adopt an HOI detector~\cite{hand} for validation. As demonstrated by our experiment results, these vision-only models without any audio information provided outperform existing models on both AVSL benchmarks. This further validates our hypothesis about visual biases in existing benchmarks. Beyond initial results, we also provide in-depth discussion with qualitative results and suggestions for further benchmark improvements. We hope our findings can shed some light on further exploration of the AVSL task and more general audio-visual learning tasks.



%% file: Sections/02_RelatedWork.tex
\section{Related Works}
\label{sec:relatedwork}

Learning to localize sound sources in videos, a task known as Audio-Visual Source Localization (AVSL), has attracted considerable research interest. Early works focus on shallow probabilistic models~\cite{probabilistic1,probabilistic2,probabilistic3} or Canonical Correlation Analysis (CCA)~\cite{cca}. With the development of deep neural networks and algorithms, self-supervised~\cite{mixture,egoloc,easyway,hardway1,hardway2,SSPL,SLAVC,SSL-TIE,lin2023unsupervised,fnac}, weakly-supervised~\cite{STM,AVGN}, and semi-supervised methods~\cite{dmt,learningtolocalsound,AVLlearning1} have been widely explored over time. For example, Sun et al.~\cite{fnac} addresses the issue of false negative samples in self-supervised contrastive learning; Guo et al.~\cite{dmt} explores a semi-supervised setting, leveraging a small amount of supervised training data to enhance audio-visual localization. In contrast to methodologies, AVSL benchmarks have received comparatively less research attention. Existing efforts on AVSL benchmarks typically fall into two categories: benchmarks focusing on videos collected from daily-life videos from the third-person view~\cite{learningtolocalsound,AVLlearning1,hardway1,hardway2}, and benchmarks focusing on egocentric videos covering human activities from the first-person view~\cite{egoloc}. In this paper, we focus on two representative benchmarks, third-person view VGG-SS~\cite{hardway1,hardway2}, and first-person view Epic-Sounding-Object~\cite{egoloc}.

%% file: Sections/03_VGG-SS.tex
\section{VGG-SS: Sounding Bias from Visual Common Sense}
\label{sec:vgg}

VGG-SS\cite{hardway1,hardway2} is an AVSL benchmark that covers a wide range of daily scenes from the third-person view. It is derived from the large-scale video dataset VGG-Sound\cite{VGGSound} and includes 5,158 ten-second video clips with audio. Each clip is annotated with bounding boxes on the middle frame for the AVSL task. In this section, we investigate the visual biases observed in VGG-SS.

\subsection{Observation and Analysis}

Ideally, test data should evaluate how well a model can identify the objects associated with a given sound among multiple possible candidates. However, based on our observations of some examples in VGG-SS, the videos typically contain only simple scenes and objects, such as a person playing an instrument or a train passing by, offering a limited number of potential candidates for sound attribution. Furthermore, with basic commonsense knowledge---such as instruments being more likely to make sounds than furniture, and moving objects being more likely to produce sounds than static ones---the sounding source in videos can often be easily distinguished based solely on visual information. We wonder whether such visual bias exists significantly in the benchmark. 
Therefore, we conduct a simple user study first. We randomly sample 300 videos from VGG-SS dataset and ask annotators whether the ground truth sounding source can be identified given only visual information, with each video annotated once. As shown in \cref{fig:human_eval}, nearly 90\% of the videos demonstrate clearly visual cues that make sounding source easily identified even without the audio. Such results intensify our concerns about the significant visual bias present in the VGG-SS dataset, casting doubt on the benchmark's effectiveness in evaluating AVSL models. To further investigate the impact of visual bias on benchmark evaluation, we conduct experiments using models with only visual input, comparing their performance to existing models specifically designed for the ASVL task.




\begin{table}[tb]
  \caption{Model performance on VGG-SS. Gray: results from a semi-supervised model that utilize partial test set data for training.}
  \label{tab:vgg_performance}
  \centering
  \begin{tabular}{@{}lccc@{}}
    \toprule
    Method & Training Data & CIoU@0.5 & AUC \\
    \midrule
     Attention10k {\scriptsize (CVPR 2018)}~\cite{AVLlearning1,learningtolocalsound} & \multirow{7}{*}{VGG-Sound 144k} & 18.50  & 30.20 \\
     Hardway {\scriptsize (CVPR 2021)}~\cite{hardway1,hardway2} & & 34.40 & 38.20 \\
     EZVSL {\scriptsize (ECCV 2022)}~\cite{easyway} & & 34.38 & 37.70 \\
     SLAVC {\scriptsize (NeurIPS 2022)}~\cite{SLAVC} & & 37.22 & 39.46 \\
     SSPL {\scriptsize (CVPR 2022)}~\cite{SSPL} & & 33.90 & 38.00 \\
     SSL-TIE {\scriptsize (ACMMM 2022)}~\cite{SSL-TIE} & & 38.60 & 39.60 \\
     FNAC {\scriptsize (CVPR 2023)}~\cite{fnac} & & 39.50 & 39.66 \\
     \midrule
     \textcolor{gray}{DMT {\scriptsize (NeurIPS 2023)}}~\cite{dmt} & \textcolor{gray}{VGG-Sound 144k, VGG-SS-4k }& \textcolor{gray}{48.80} & \textcolor{gray}{45.76} \\ \midrule
     MiniGPT-v2 & - & \textbf{48.88} & \textbf{49.07} \\
  \bottomrule
  \end{tabular}
\end{table}

\subsection{Experiments}

\noindent \textbf{Model.} We choose MiniGPT-v2~\cite{minigptv2} as the vision-only model. Building upon large vision and language models~\cite{Vit,llama2} and pretrained with diverse vision-language tasks, it is not only proficient in general vision-language understanding and reasoning, but also capable of visual elements grounding. For grounding tasks, the model autoregressively generates a sequence of four values representing the coordinates of the top-left and bottom-right corners of the bounding box. To adapt the model to the AVSL task, we prompt the model to locate the most likely sounding source in the given video frame. Since the model occasionally fails to follow instructions accurately, we construct a pseudo training corpus by rephrasing the MSCOCO~\cite{coco} detection data to instruction data, which helps to better align the model with the desired task output format. We finetune the model on the pseudo training data for 10 epochs with LoRA~\cite{lora}.


\noindent \textbf{Baselines and Evaluation.}
We use recent models benchmarked on VGG-SS as our baselines, including Attention10k~\cite{AVLlearning1,learningtolocalsound}, Hardway~\cite{hardway1,hardway2}, EZVSL~\cite{easyway}, SLAVC~\cite{SLAVC}, SSPL~\cite{SSPL}, SSL-TIE~\cite{SSL-TIE}, and FNAC~\cite{fnac}, all of which adopt unsupervised contrastive learning frameworks for audio-visual learning. Among them, FNAC serves as the strongest baseline with false negatives suppression and true negatives enhancement. Moreover, we also include a recently proposed semi-supervised framework DMT~\cite{dmt}, which utilizes about 80\% of the VGG-SS test data for training and reports the performance evaluated on the remaining data. All results are obtained from the original papers. 
Following previous works~\cite{AVLlearning1,learningtolocalsound}, we adopt the Consensus IoU (CIoU) and AUC as metrics for model evaluation, with a cIoU threshold of 0.5.

\noindent \textbf{Results.}
\cref{tab:vgg_performance} presents the results of different models on the VGG-SS benchmark, where our vision-only MiniGPT-v2 model obviously outperforms all ASVL baselines, and even surpasses the semi-supervised model DMT trained with about 80\% of the test set data. This demonstrates that models can achieve competitive performance using only visual information, without the need for audio information input, further confirming our hypothesis about the visual bias in the VGG-SS benchmark.


%% file: Sections/04_Epic-SS.tex
\begin{table}[tb]
  \caption{Model performance on Epic-Sounding-Object.}
  \label{tab:epic_performance}
  \centering
  \begin{tabular}{@{}lccc@{}}
    \toprule
    Method & Training Data & CIoU@0.2 & AUC \\
    \midrule
     Attention10k {\scriptsize (CVPR 2018)}~\cite{AVLlearning1,learningtolocalsound} & \multirow{6}{*}{Epic-Kitchen} & 7.12 & 6.42 \\
     STM {\scriptsize (BMVC 2021)}~\cite{STM} & & 12.10 & 8.87 \\
     Hardway {\scriptsize (CVPR 2021)}~\cite{hardway1,hardway2} & & 24.51 & 13.38 \\
     Mix {\scriptsize (CVPR 2022)}~\cite{mixture} & & 26.01 & 15.39 \\
     SSPL {\scriptsize (CVPR 2022)}~\cite{SSPL} & & 13.62 & 9.56 \\
     EgoLoc {\scriptsize (CVPR 2023)} ~\cite{egoloc} & & 38.71 & 18.38 \\ \midrule
     HOID & DOH100k~\cite{hand} & 32.53 & 32.60 \\
     HOID & DOH100k~\cite{hand}, Ego~\cite{egodata1,egodata2,Epic1,Epic2} & \textbf{43.54} & \textbf{43.30} \\
  \bottomrule
  \end{tabular}
\end{table}

\section{Epic-Sounding-Object: Sounding Bias from Hand-Object Interaction}
\label{sec:epic}
Epic-Sounding-Object~\cite{egoloc} is a first-person AVSL benchmark mainly capturing scenes in kitchens. It is derived from the large-scale egocentric video dataset Epic-Kitchens~\cite{Epic1,Epic2}. It contains 3,172 video clips, most of which have an average duration of less than 3 seconds. Each clip is annotated with bounding boxes for the AVSL task on the middle frames. This section investigates the visual biases in Epic-Sounding-Object benchmark.

\begin{figure}[tb]
  \centering
  \includegraphics[width=0.8\linewidth]{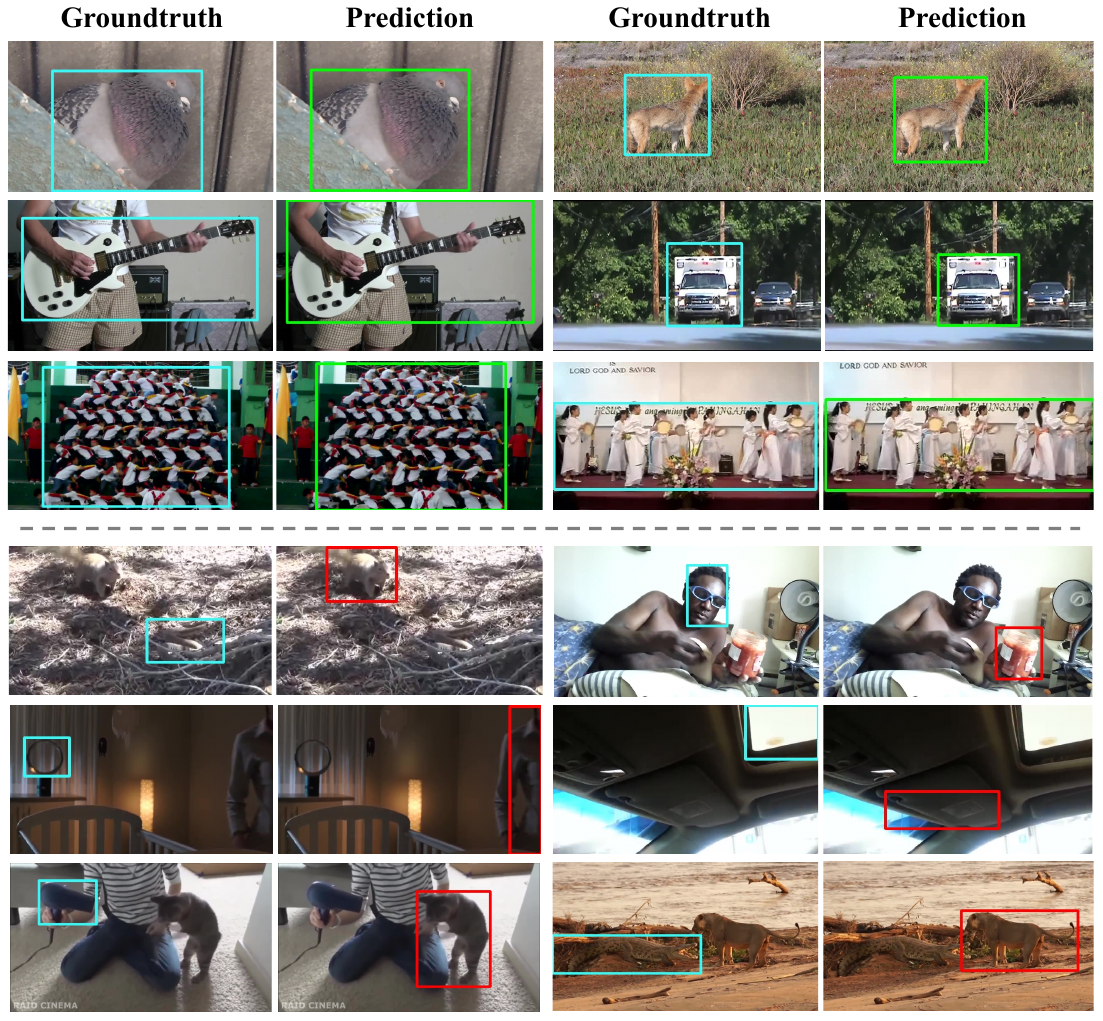}
  \caption{ \textbf{Qualitative results on VGG-SS. Top: well-performed cases; Bottom: failure cases.}}
  \label{fig:vgg_good}
\end{figure}

\begin{figure}[h!]
  \centering
  \includegraphics[width=0.8\linewidth]{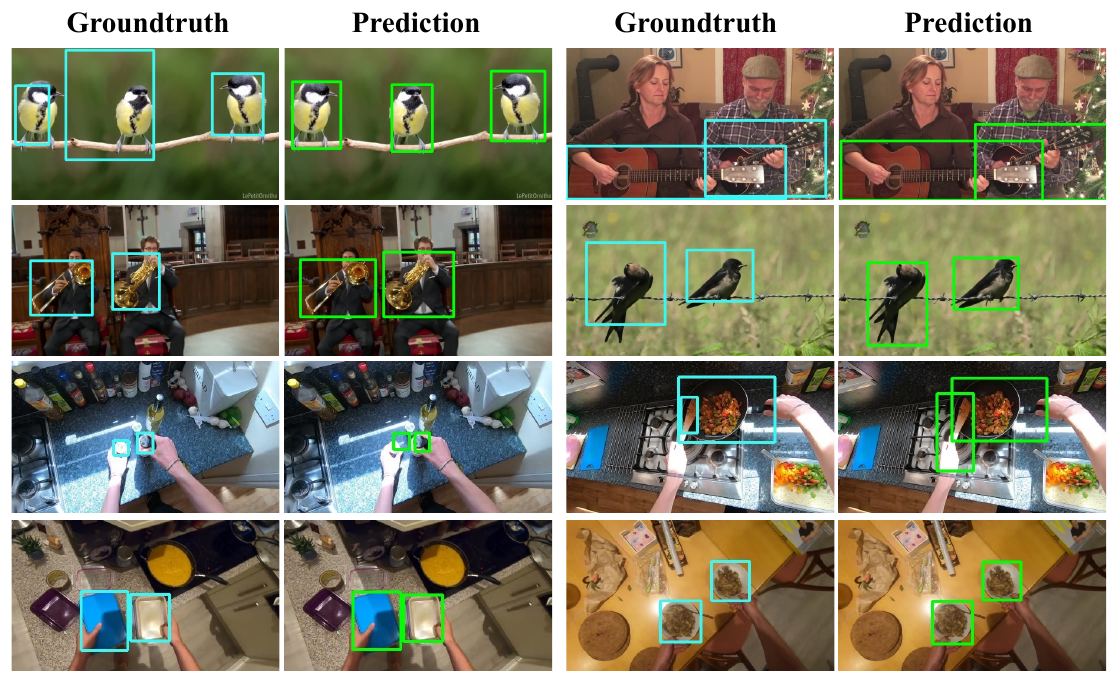}
  \caption{ \textbf{Qualitative results on multi-source audio-visual localization data.}}
  \label{fig:multi_source}
\end{figure}

\subsection{Observation and Analysis}

In the egocentric videos in kitchens, sounding objects can often be identified using solely visual information as well. Sounds in a kitchen typically originate from human activities, such as chopping vegetables or washing dishes, while static objects are more likely to remain silent. Moreover, people familiar with cooking can easily discern which kitchen utensils are more likely to produce sounds. As a result, locating the sound source in kitchen videos is not a difficult task, even without audio information. Our manual review of over 300 random samples further confirms this point---in more than 90\% of the cases, the sound source could be accurately inferred (consistent with the ground truth) using only visual information.

However, kitchen scenes pose greater complexity compared to those in VGG-SS, featuring more and diverse objects that demand fine-grained visual understanding from models and introduce more candidates for distinction. Moreover, unlike the naturally sounding objects in VGG-SS, such as musical instruments and animals, the sound sources in Epic-Sounding-Objects, such as object collisions or boiling water, are less intuitive and straightforward, presenting more challenges for simple visual commonsense reasoning to identify. Consequently, reasoning sound sources from kitchen videos generally presents greater challenges for models. However, we find that Epic-Sounding-Objects primarily focus on human actions, especially hand activities, suggesting that simply locating the interactions between hands and objects can likely pinpoint the correct sound source. Therefore, in this section, we utilize a Human-Object Interaction (HOI) detector~\cite{hand} to analyze the visual bias within this benchmark.



\subsection{Experiments}

\noindent \textbf{Model.} We employ Hand Object Interaction Detector (HOID)~\cite{hand}, an HOI detection model with a ResNet-101~\cite{he2016deep} backbone pretrained on ImageNet~\cite{deng2009imagenet}, for AVSL benchmark validation. We test two versions of HOID, a basic version pretrained on a general domain dataset DOH100k~\cite{hand}, and an enhanced version additionally pretrained with 42K egocentric data~\cite{egodata1,egodata2,Epic1,Epic2} to better generalize to egocentric HOI detection.


\noindent \textbf{Baselines and Evaluation.} We compare HOID with several recent baselines on Epic-Sounding-Object, including Attention10k~\cite{AVLlearning1,learningtolocalsound}, STM~\cite{STM}, Hardway~\cite{hardway1,hardway2}, Mix~\cite{mixture} and EgoLoc~\cite{egoloc}.
Following previous works~\cite{egoloc}, we use CIoU@0.2 and AUC as metrics on Epic-Sounding-Object.


\noindent\textbf{Results.} As shown in \cref{tab:epic_performance}, both HOID models perform well on the Epic-Sounding-Object benchmark, and the enhanced version trained with egocentric data outperforms all baselines by a large margin. This demonstrates that simply identifying hand-object interactions can often correctly localize the sound source in Epic-Sounding-Object benchmark, revealing an obvious visual bias that diminishes the benchmark's suitability for validating AVSL models.


\begin{figure}[tb]
  \centering
  \includegraphics[width=0.8\linewidth]{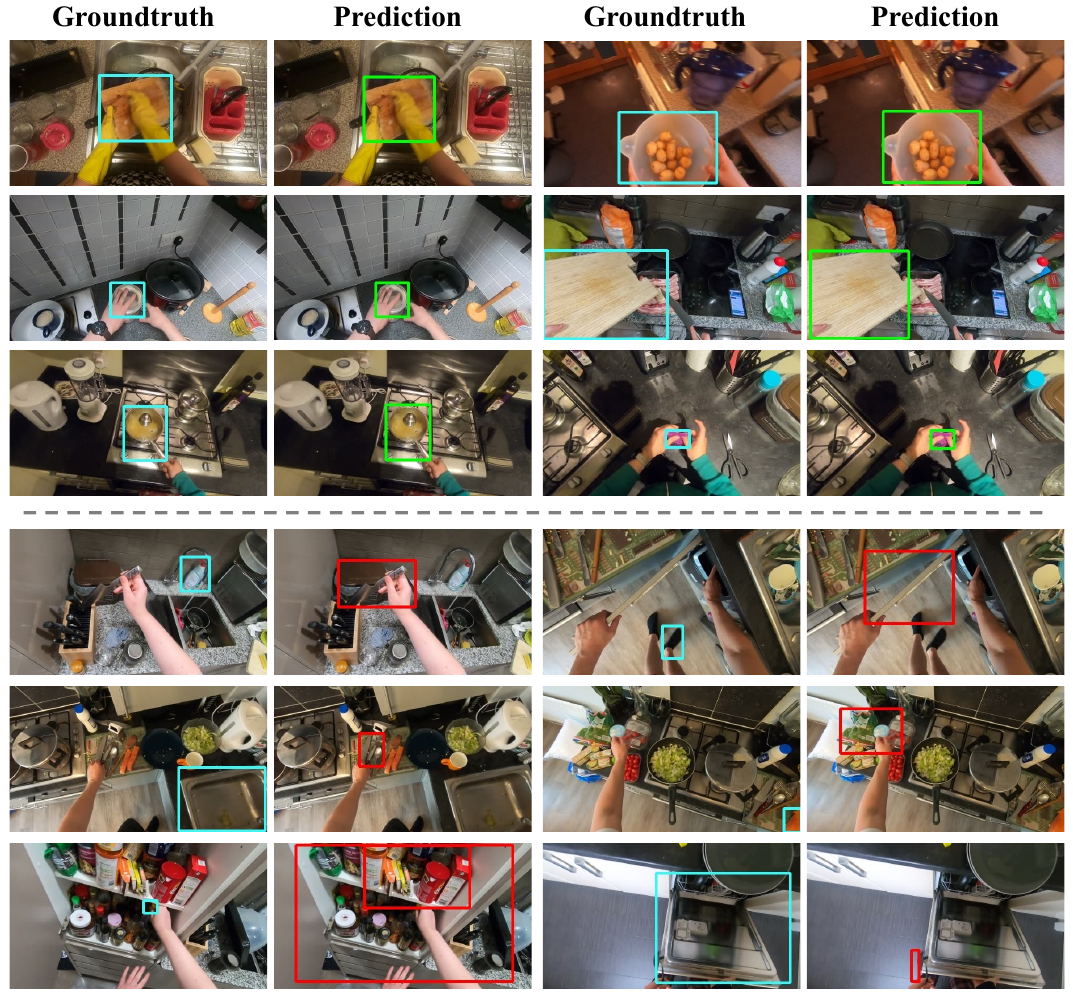}
  \caption{ \textbf{Qualitative results on Epic-Sounding-Object. Top: well-performed cases; Bottom: failure cases.}}
  \label{fig:epic_good}
\end{figure}

%% file: Sections/05_Discussion.tex
\section{Discussion}
\label{sec:discussion}


This section presents our observations on VGG-SS and Epic-Sounding-Object, along with qualitative results. Based on these observations, we propose suggestions for future improvements to the benchmarks.


\noindent \textbf{VGG-SS.} 
We provide some qualitative results of MiniGPT-v2 on VGG-SS. As shown in \cref{fig:vgg_good} (Top), given only video frames, the model accurately identifies the sound sources in these videos, such as animals and musical instruments. It is reasonable to expect that MiniGPT-v2 can easily pinpoint these targets, as they are typically recognized as naturally sounding objects and serve as the main subjects in the video. Another interesting finding is that MiniGPT-v2 exhibits potential in multi-source localization as well, as shown in \cref{fig:multi_source}.
However, \cref{fig:vgg_good} (Bottom) illustrates some failure cases where the model struggles with only video frames as input: (1) the reasoning of sound source relies on the video motion, which cannot be captured from a single frame (Row 1); (2) the target objects are difficult to distinguish or easily confused (Row 2); and (3) multiple potential sounding objects present simultaneously (Row 3). While the first type of failure case can be addressed by utilizing a video model to handle motion information, the remaining cases require audio information for the model to accurately determine the true sounding source in a video.

\noindent \textbf{Epic-Sounding-Object.} 
As most of the sound sources in Epic-Sounding-Object are directly linked to interactions between hands and objects, a simple HOI detection method achieves precise localization in many cases, as illustrated in \cref{fig:epic_good} (Top), highlighting the visual bias present in this benchmark. However, as shown in \cref{fig:epic_good} (Bottom), for data instances without such visual biases, i.e., the sounding objects are not in contact with the hands, the HOI detector often fails to locate the target. In these cases, audio information typically becomes crucial for sounding object localization.

\noindent \textbf{Future Benchmark Improvement.} By hacking the existing benchmarks with vision-only models, we demonstrate that these benchmarks are plagued by significant visual biases that hinder their effectiveness in evaluating audio-visual models. As observed from the qualitative cases above, successfully predicted cases typically exhibit more visual biases than failure ones. This suggests that a straightforward strategy to mitigate visual biases and enhance the benchmarks is to filter out the data that can be easily hacked by vision-only models. We leave this for further investigation.


%% file: Sections/06_Conclusion.tex
\section{Conclusion}
\label{sec:conclusion}
In this paper, we revealed some significant visual biases in Audio-Visual Source Localization (AVSL) benchmarks. We first demonstrated that visual information alone can often suffice for accurately localizing the sounding objects within visual scenes through a simple user study. We further validated the presence of such visual biases in two representative AVSL benchmarks through experiments using vision-only models without audio input like MiniGPT-v2 and a hand-object interaction detector (HOID), which surpassed all other AVSL models. These findings suggest a need for benchmark refinement to better support audio-visual learning and other audio-visual downstream tasks. 